\documentclass[letter]{aa}

\usepackage{graphicx}
\usepackage{txfonts}
\usepackage{hyperref}
\hypersetup{
    pdftitle={Blazar 1424+240},
    pdfauthor={Yuri Kovalev},     
    colorlinks=true,       
    linkcolor=blue,
    filecolor=magenta,      
    urlcolor=violet,
    citecolor=teal,        
    pdfborder={0 0 0}
}
\usepackage{orcidlink}
\usepackage{xspace}

\graphicspath{{./}{figs/}}

\newcommand{\pks}{PKS\,1424+240\xspace}

\begin{document} 

\title{
Looking into the jet cone of the neutrino-associated very high energy blazar PKS\,1424+240
}
\titlerunning{
The high energy multi-messenger blazar \pks
}

\author{
Y.~Y.~Kovalev\inst{1}\orcidlink{0000-0001-9303-3263}\thanks{\email{yykovalev@gmail.com}}
\and
A.~B.~Pushkarev\inst{2,3}\orcidlink{0000-0002-9702-2307}
\and
J.~L.~Gómez\inst{4}\orcidlink{0000-0003-4190-7613}
\and
D.~C.~Homan\inst{5}\orcidlink{0000-0002-4431-0890}
\and
M.~L.~Lister\inst{6}\orcidlink{0000-0003-1315-3412}
\and
J.~D.~Livingston\inst{1}\orcidlink{0000-0002-4090-8000}
\and
I.~N.~Pashchenko\inst{3}\orcidlink{0000-0002-9404-7023}
\and
A.~V.~Plavin\inst{7}\orcidlink{0000-0003-2914-8554}
\and
T.~Savolainen\inst{8,9,1}\orcidlink{0000-0001-6214-1085}
\and
S.~V.~Troitsky\inst{10,11}\orcidlink{0000-0001-6917-6600}
}

\institute{
Max Planck Institute for Radio Astronomy, Auf dem Hügel 69, D--53121 Bonn, Germany
\and
Crimean Astrophysical Observatory, 298409 Nauchny, Crimea
\and
Lebedev Physical Institute of the Russian Academy of Sciences, Leninsky prospekt 53, 119991 Moscow, Russia
\and
Instituto de Astrofísica de Andalucía-CSIC, Glorieta de la Astronomía s/n, 18008 Granada, Spain
\and
Department of Physics and Astronomy, Denison University, Granville, OH 43023, USA
\and
Department of Physics and Astronomy, Purdue University, 525 Northwestern Avenue, West Lafayette, IN 47907, USA
\and
Black Hole Initiative, Harvard University, 20 Garden St, Cambridge, MA 02138, USA
\and
Aalto University Mets{\"a}hovi Radio Observatory, Mets{\"a}hovintie 114, FI-02540 Kylm{\"a}l{\"a}, Finland
\and
Aalto University Department of Electronics and Nanoengineering, PL 15500, FI-00076 Aalto, Finland
\and
Institute for Nuclear Research,
60th October Anniversary Prospect 7a, Moscow 117312, Russia
\and
Physics Department, Lomonosov Moscow State University, 1-2 Leninskie Gory, Moscow 119991, Russia
}

\date{Received 6 May, 2025; accepted 7 July, 2025}

\abstract
{
The acceleration process of massive particles as well as the production of very high energy (VHE) photons and neutrinos remains a fundamental challenge in astrophysics.
}
{
We investigate the parsec-scale jet structure and magnetic field of the blazar \pks, that has been selected on the basis of strong VHE gamma-ray emission and identified with one of the highest peaks in the IceCube 9-year neutrino sky.
}
{
We analyze 15~GHz VLBA observations of this BL~Lac object, stacking 42 polarization-sensitive images collected in 2009--2025 to enhance the signal and reveal persistent parsec-scale structure.
}
{
Our observations uncover a rare scenario. The object is viewed inside the jet cone, very close to the axis of its relativistic jet, with a viewing angle of $<0.6^\circ$. This effectively maximizes Doppler boosting to values $\sim30$ and enhances both electromagnetic and neutrino emission in the observer’s direction. Based on polarimetric observations, we unambiguously detect a net toroidal component of the jet’s magnetic field, indicating a current carrying jet flowing almost directly towards our line of sight.
}
{
Blazars with very small jet viewing angles offer a solution to the longstanding mismatch between Doppler factors inferred from low VLBI apparent jet speed and those derived from VHE observations~--- the so-called ``Doppler factor crisis''.
We show that relativistic beaming plays the critical role in the gamma-ray and neutrino emission of blazars, with direct implications for models of their multi-messenger emission.
}


\maketitle
\nolinenumbers

\section{Introduction}

Understanding how the universe produces photons and neutrinos at the highest observable energies is of broad interest, as it directly relates to fundamental physics that can only be explored through astronomy. 
Active galactic nuclei (AGN) are strong emitters of very high energy (VHE) photons \citep[e.g.,][]{2007ApJ...669..862A,2007ApJ...664L..71A} and have been proposed as efficient proton accelerators \citep[starting with][]{BerezinskyNeutrino77}, capable of producing neutrinos at TeV--PeV and beyond. 
Their observed properties are strongly influenced by relativistic beaming effects that also introduce significant observational biases. For example, flux-density-limited samples of radio Very Long Baseline Interferometry (VLBI) selected AGN are dominated by distant beamed blazars with a median redshift of $z\sim1$ \citep{2025ApJS..276...38P}. 
However, the role of beaming at very high energies ($\gtrsim 100$~GeV) is not firmly established. Indeed, while fast variability of VHE gamma-rays implies high values of the Doppler factor $\delta$, the lack of strong apparent superluminal motions in VLBI observed jets suggests the opposite.
This is known as the ``Doppler factor crisis'' \citep[see, e.g.,][]{2007ApJ...669..862A,2007ApJ...664L..71A,2018ApJ...853...68P}.
Breakthroughs in this area have been rare over the past 35 years, with the Doppler factor crisis posing a significant barrier to progress.
The neutrino emission from distant blazars is expected to be beamed, with proton acceleration occurring preferentially along the jet direction \citep[e.g.,][]{0506-scienceGamma,nubeam}. In contrast, weaker AGN may emit neutrinos 
\citep{IC1068} without any preferred direction.

\begin{figure}
    \centering
    \includegraphics[width=1.00\linewidth,trim=0.3cm 0.4cm 0.3cm 0cm]{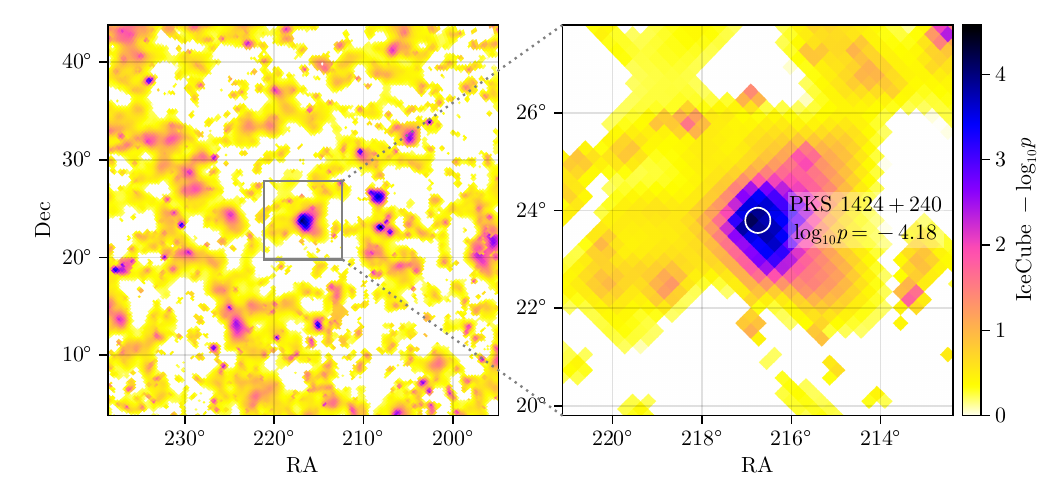}
    \caption{The IceCube skymap cutout surrounding the radio position of the \pks blazar. The color scale represents the local p-value from the 9-year maximum likelihood analysis performed by \citet{IC1068}. 
    }
    \label{f:icmap}
\end{figure}

With a redshift $z=0.605$ \citep{2017ApJ...837..144P}, the BL~Lac object \pks (common name \object{OQ\,240}) is the most distant blazar detected in quiescent VHE \citep[e.g.,][]{2014A&A...567A.135A,Padovani1424}.
The all-sky analysis performed by \citet{IC1068} has identified it as the possible source of the second-highest neutrino excess in the predefined source list in the Northern sky (see \autoref{f:icmap}).
The best-fit neutrino spectrum for \pks is soft with a spectral index of $\gamma = 3.5$ \citep{IC1068}. The IceCube signal is dominated by neutrinos with relatively low, $\lesssim 10$~TeV, energies. The time-integrated IceCube analysis is predominantly sensitive to persistent emission. Together with off-flare VHE detections in gamma-rays, this motivates the study of time-independent properties of \pks. Here, we address the question, which general property of \pks may be responsible for its exceptional VHE photon and neutrino fluxes. 

We use radio VLBI, which is the only technique to date that is able to directly probe the sites of neutrino and gamma-ray production, resolving structures within several parsecs from the central engine. For the redshift of \pks, the projected angular scale of 1~mas corresponds to 6.70~pc, assuming the \citet{Komatsu09} cosmological parameters.
A quick examination of \pks has revealed that its jet has an exceptionally large apparent opening angle $\varphi_\mathrm{app}$. 
\citet{MOJAVE_XIV} found $\varphi_\mathrm{app} = 65\degr$, among the most extreme values observed for blazars. Geometrically, this indicates a small viewing angle $\theta$ to the line of sight and, consequently, strong Doppler boosting. To further investigate the source, we updated its stacked 15~GHz image by the Very Long Baseline Array (VLBA) using all available data and increasing the number of stacked epochs by about $\times4$. The results of our serendipitous findings are presented in this paper.

\section{15~GHz VLBA polarization imaging}
\label{s:imaging}

Stokes I and polarization observation of \pks in the MOJAVE program\footnote{\url{https://www.cv.nrao.edu/MOJAVE/}} \citep[\mbox{MOJAVE:} Monitoring Of Jets in Active galactic nuclei with VLBA Experiments, see][and references therein]{2018ApJS..234...12L} began in May~2009 and we report results through January~2025.
%
%
%
To increase the dynamic range of a restored map, a set of single-epoch images was combined and averaged, i.e., stacked, by aligning them using the VLBI core position derived from visibility model fitting \citep[e.g.,][]{2021ApJ...923...30L}. Stacking not only improves the sensitivity and dynamic range but also effectively reconstructs more complete source morphology, particularly in the low-brightness areas. It can also reveal the whole jet channel, not just regions of enhanced radiation at a given epoch. Image stacking is beneficial for studies of jets in Stokes I and polarized emission \citep[e.g.][]{MOJAVE_XIV,MOJAVE_XX,2020MNRAS.495.3576K,MOJAVE_XXI}.
%

\begin{figure}[t]
\centering
\includegraphics[width=1.00\columnwidth]{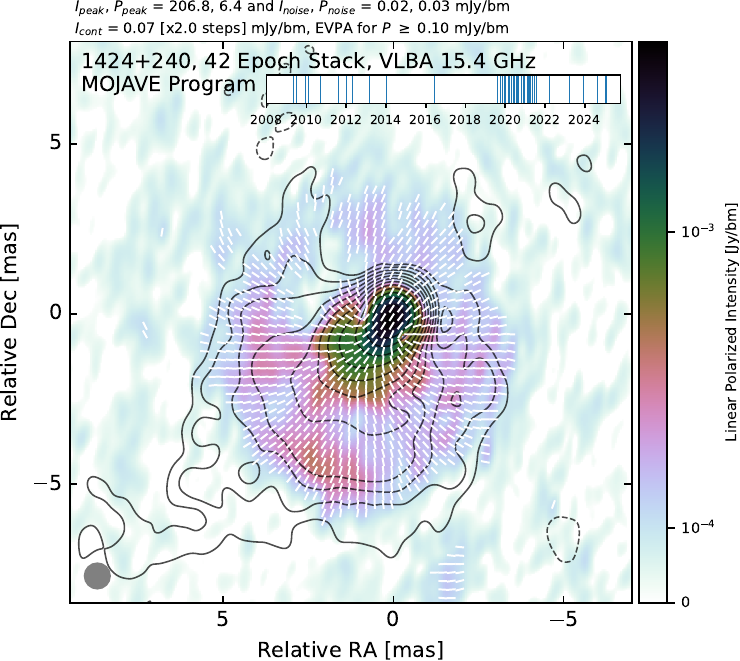}
\caption{VLBA stacked image for the blazar \pks at 15~GHz. Stokes I is shown by contours, the first contour corresponds to $3\times$ the image rms level. Linear polarization intensity is presented by color, directions of electric vector position angle (EVPA)~--- by sticks. The observation dates of the 42 images used in the stacked image are indicated in the inset. A circular restoring beam is shown at the full width at half maximum (FWHM, 0.8~mas) level in the bottom left corner.
FITS files of the stacked Stokes I, Q, and U images are available at the CDS.
We also supply an animation of Stokes I cumulative stacking (available online).
}
\label{f:1424stack}
\end{figure}

We made stacked maps of \pks in Stokes I, Q, and U (\autoref{f:1424stack}), using 42 available epochs that span approximately 15 years, following a procedure described in \citet{MOJAVE_XX}. For deep CLEANing in Stokes I, Q, and U, we applied the algorithm with the residual entropy-based stopping criterion developed by \citet{Homan24}. The linear polarization image has been corrected for Ricean bias \citep{Wardle74}. We also applied a debiasing procedure to the CLEAN images to effectively suppress the sidelobe artifacts from deconvolution \citep[Appendix B in][]{MOJAVE_XX}. 

This debiasing procedure has significantly reduced imaging artifacts, and we argue that the weak structures observed around the core of \pks in the stacked images (\autoref{f:1424stack}) represent genuine reconstructed jet emission.
The northern declination of \pks ensures robust VLBA $uv$-coverage with no significant gaps.
The weak polarized emission stacks coherently over multiple epochs, and is not expected to result from residual polarization leakage effects.
The reduced levels of detectable 15~GHz Stokes I emission in the North-West direction next to the bright core component possibly results from residual effects of self-calibration which only included emission to the South-East in the self-calibration model in most individual epochs.
In order to check for consistency the \autoref{f:1424stack} Stokes~I image, we have processed archival 1.6~GHz VLBA data for \pks, see \autoref{s:l-band}.

\section{Looking into the jet cone}
\label{s:jet_cone}

\subsection{Observational signatures and magnetic field}
\label{s:obs_sign}

\begin{figure}[t]
\centering
\includegraphics[width=1.00\columnwidth,trim=0.2cm 0.1cm 0.2cm 0.2cm]{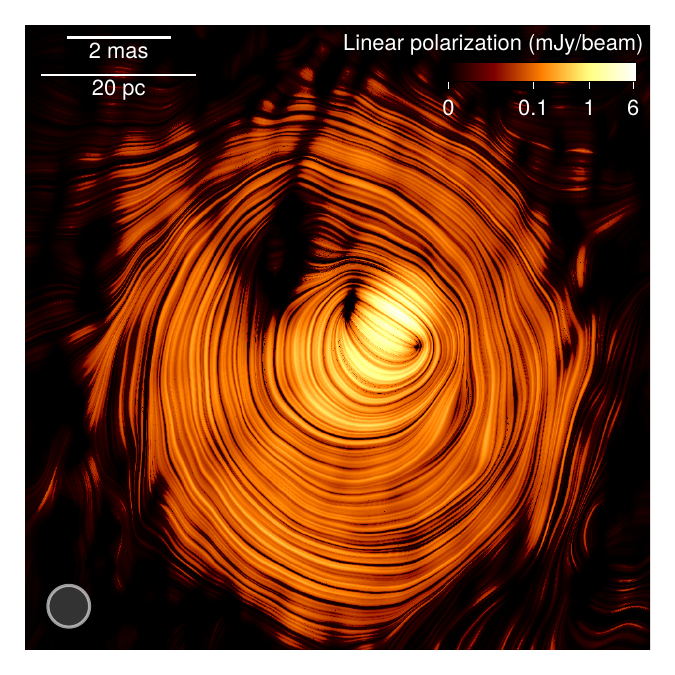}
\caption{
The Eye of Sauron.
Stacked VLBA linear polarization image with magnetic field direction  incorporated by the Linear Integral Convolution.
The circular restoring beam is shown at the FWHM level of 0.8~mas in the bottom left corner.
}
\label{f:sauron_eye}
\end{figure}

We have collected striking observational signatures of a jet that is viewed inside its cone.
The stacked Stokes I and linear polarization images reveal a structure in the southeast direction, while significant emission is also present in all other directions around the core region in Stokes I and linear polarization (\autoref{f:1424stack}).
This is extended even to deprojected kiloparsec scales, see \autoref{f:L_image}.
Furthermore, the stacked polarization image exhibits a remarkably uniform distribution of EVPAs; it presents a pattern of ``diverging rays'', which start from the core.
Rotation measure 15/24/43~GHz VLBA results are presented in \autoref{s:RM}.
The low rotation measure and its minimal variation suggest that the jet is propagating toward the observer, reducing the density of the interstellar medium along its path.
The low jet speeds, observed by VLBI (\autoref{s:jet_parameters}), are consistent with a small jet viewing angle, $\theta$.

We reconstruct the magnetic field structure, implementing a $90^\circ$ EVPA rotation across the entire map.
This assumes that RM as well as relativistic effects do not affect the observed projected magnetic field direction significantly, see \autoref{s:RM} and \citet{2005MNRAS.360..869L}.
The results are shown in \autoref{f:sauron_eye}, which (after J.~R.~R.~Tolkien) we call `the Eye of Sauron' due to its striking nature. 
Linear Integral Convolution \citep[LIC;][]{LIC} is applied to the polarized intensity to visualize the morphology of magnetic field lines by locally blurring intensity variations along the magnetic field direction, effectively tracing its structure.
The image reveals a significant toroidal magnetic field component of a current-carrying jet, which flows almost directly toward Earth. This is consistent with a widely discussed overall helical magnetic field structure \citep{2019ARA&A..57..467B}.

\subsection{Basic jet parameters}
\label{s:jet_parameters}

\citet{2021ApJ...923...30L} have measured apparent kinematics of two robust components in the emission of \pks. Their apparent speed, measured in units of the speed of light, is $\beta_\mathrm{app,1}=2.83\pm0.89$ and $\beta_\mathrm{app,2}=1.91\pm0.18$. These moderate values are consistent with other measurements for the same source \citep[e.g.,][]{Padovani1424,Kun1424}. 
A median intrinsic full opening angle of the jet was estimated by \citet{MOJAVE_XIV} to be $\alpha_\mathrm{int}=1.2^\circ$ for \textit{Fermi}-detected AGN in MOJAVE sample. We assume this value for \pks.
Looking into the jet cone condition is defined by the requirement for the jet viewing angle $\theta<\alpha_\mathrm{int}/2$. 
This requires the Doppler factor to reach its maximum possible value of twice the Lorentz factor, $\delta\approx2\Gamma$.

We apply a simplified assumption that the jet is observed at a half of its half opening angle, $\theta=\alpha_\mathrm{int}/4$. This is in agreement with simulations by Pashchenko et al.~(in prep.).
We also assume a relation between $\alpha_\mathrm{int}$ and $\Gamma$, as deduced observationally by \citet{MOJAVE_XIV} and predicted by hydrodynamical and magnetic acceleration models  of relativistic jets
\citep{BK79,2007MNRAS.380...51K,2013A&A...558A.144C}.
As a result, we find that the following set of parameters for a basic relativistic jet model are consistent with the above values and considerations:
$\Gamma = 16$ (plasma flow speed $\beta = 0.998$),
$\alpha_\mathrm{int} = 1.2^\circ$,
$\theta = 0.3^\circ$,
$\beta_\mathrm{app} = 2.8$,
and Doppler factor $\delta = 32$.
The Lorentz factor is within its typical values from MOJAVE population modeling \citep{2019ApJ...874...43L}, as expected from the chosen conservative parameter assumptions.
The Doppler factor is three times higher than the median values for the MOJAVE sample \citep{2021ApJ...923...30L,2021ApJ...923...67H}, which agrees with recent finding of \cite{nubeam} for neutrino-selected blazars. Note that $\Gamma$ and $\delta$ estimates are not highly sensitive to the exact value of $\theta$, as long as the line of sight remains within the jet cone.
Assuming a jet viewing angle in the range $0.1^\circ < \theta < 1.0^\circ$ and using the measured $\beta_\mathrm{app}$ values, the Doppler factors lie within the range $52>\delta>16$.
Although the geometric probability of a near-zero viewing angle is low, the fraction of such jets in VLBI flux density-selected samples has been shown to be on the order of a few percent due to Doppler bias \citep{2019ApJ...874...43L}.

Due to strong projection effects and synchrotron opacity, the jet is expected to become visible many parsecs downstream from its true base. As a result, its observed radio core appears less active and variable than typically expected for highly boosted blazars.
Moreover, the core and jet emission are superimposed, resulting in a significant underestimation of the intrinsic core brightness temperature, which in turn leads to an incorrect estimation of jet parameters in \citet{2021ApJ...923...67H}.
A moderate flare occurred in the core of \pks during 2019–2020; the core flux density has reached 0.27~Jy, while the median value over all epochs is 0.18~Jy. It did not affect these estimates significantly.

\section{Emission of high energy photons and neutrinos: relativistic beaming}

The ``Doppler factor crisis'' outlines the contradiction between a short time scale of the blazar high-energy emission variability and low estimates of the Doppler factor for VHE blazars \citep[e.g.,][]{2010ApJ...722..197L}. 
Apparent VLBI speed for VHE blazars is observed to be typically less than 2--3$c$ \citep{2018ApJ...853...68P}.
We propose a partial solution of this problem, linking low apparent speed with a very small jet viewing angle.
Our observations suggest that the Doppler factors inferred from VHE and radio observations may be consistent, not requiring different sites of these emissions. 

We note that the explanation of the VHE gamma-ray flux within a leptohadronic model of \citet{Cerruti1424} requires the Doppler factor of $\sim 30$, consistent with our observations. The expected associated neutrino flux is of order of that inferred from IceCube observations. The neutrino production in the blazar radio core may proceed through the $p\gamma$ mechanism, adopted by \citet{neutradio2} and \citet{Polina} to explain TeV to PeV neutrino production. Energetic protons, required for the neutrino and, in the hadronic scenarios, gamma-ray production, may be accelerated either in the vicinity of the central black hole \citep{Ptitsyna-gap} or at the border surface between the fast spine and the slower sheath in the jet \citep[e.g.,][]{neutrino-from-spine-sheath}. In any case, 
the specific geometry of the jet in \pks, which is closely aligned to the line of sight, provides an additional Doppler enhancement of its multi-wavelength and multi-messenger fluxes. 
This boosting makes the object persistently bright, maintaining a high average flux~--- placing it among the top 1\% of gamma-ray sources \citep{4FGL-DR4}, and the brightest blazar in terms of high-energy neutrino emission \citep{IC1068}.
Future observations of this and other VHE emitting blazars are needed to pave the way to a quantitative model of neutrino production in jets, and to better understand the role of beaming in the gamma-ray emission.

\section{Summary}
\label{s:summary}

Supermassive black holes at the heart of distant active galaxies are believed to be among the most powerful particle accelerators in the Universe. They produce VHE photons and serve as possible sources of cosmic neutrinos. The mechanism responsible for accelerating massive particles to relativistic energies remains an open question in modern astrophysics, with powerful extragalactic jets being a key candidate. A longstanding discrepancy~--- known as the Doppler factor crisis~--- arises from the mismatch between low apparent plasma speeds observed with VLBI and the highly relativistic jet required to explain the emission.
We use 16~years of 2~cm VLBI observations to uncover properties of the parsec-scale jet of the blazar \pks, which is exceptional in its VHE gamma-ray and probable neutrino emissions.
A significant toroidal magnetic field is detected by a clear signature in linear polarization, revealing a current-carrying jet flowing almost directly toward us. We determine that the jet is observed within its cone, at a viewing angle below half a degree. This geometry maximizes relativistic boosting while minimizing apparent speed, thereby resolving the Doppler factor crisis and suggesting  extreme relativistic beaming for this VHE photon and neutrino-emitting blazar. Our findings for \pks pave the way towards better understanding of high-energy photon and neutrino emission from blazar jets.

Observing jets within their plasma cone provides a unique opportunity to determine their intrinsic parameters and reconstruct the projected magnetic field structure. Moreover, such jets may represent a distinct class of AGN with both enhanced VHE gamma-ray and neutrino emission.
From Monte Carlo simulations of the complete, VLBI-selected MOJAVE 1.5JyQC sample, we estimate that only a few percent of the jets are viewed within a degree to our line of sight \citep{2019ApJ...874...43L}.
The results of a full MOJAVE sample analysis will be presented separately by Pushkarev et al.\ (in prep.).

\section{Data availability}

Stacked Stokes~I, Q, U 15~GHz FITS images (\autoref{f:1424stack}) as well as Stokes~I 1.6~GHz FITS image (\autoref{f:L_image}) are only available in electronic form at the CDS via anonymous ftp to cdsarc.u-strasbg.fr (130.79.128.5) or via \url{http://cdsweb.u-strasbg.fr/cgi-bin/qcat?J/A+A/}.

\begin{acknowledgements}

We thank Andrei Lobanov, Yannis Liodakis, and Tigran Arshakian for productive discussions and comments on the manuscript.
This work is part of the MuSES project which has received funding from the European Research Council (ERC) under the European Union’s Horizon 2020 Research and Innovation Programme (grant agreement No~101142396).
The rotation measure analysis by JDL is supported by the M2FINDERS project, which has received funding from the ERC under the European Union’s Horizon 2020 research and innovation programme (grant agreement No~101018682).
AVP is a postdoctoral fellow at
the Black Hole Initiative, which is funded by grants from the John Templeton Foundation (grants 60477, 61479, 62286) and the Gordon and Betty Moore Foundation (grant GBMF-8273). 
The work of SVT is supported in the framework of the State project ``Science'' by the Ministry of Science and Higher Education of the Russian Federation under the contract 075-15-2024-541.
The opinions expressed in this work are those of the authors and do not necessarily reflect the views of these Foundations.

The National Radio Astronomy Observatory is a facility of the National Science Foundation operated under cooperative agreement by Associated Universities, Inc.
This research has made use of the NASA/IPAC Extragalactic Database, which is funded by the National Aeronautics and Space Administration and operated by the California Institute of Technology.
This research made use of the data from the MOJAVE database maintained by the MOJAVE team \citep{2018ApJS..234...12L}.

\end{acknowledgements}

\bibliographystyle{aa}
\bibliography{1424}

\begin{appendix}
\section{Supporting observational material}
\subsection{VLBA data at 1.6~GHz}
\label{s:l-band}

\begin{figure}
\centering
\includegraphics[width=1.00\columnwidth]{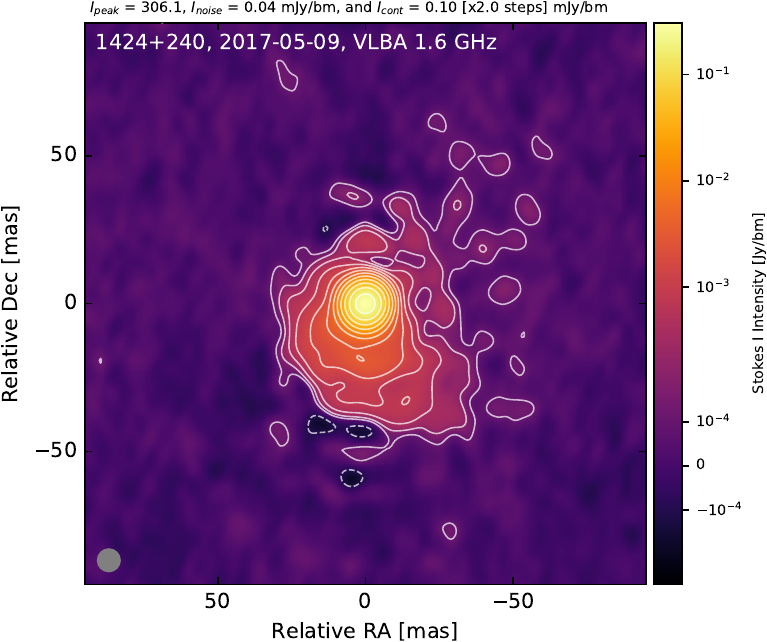}
\caption{1.6~GHz VLBA image of \pks, epoch 9~May~2017. 
A circular restoring beam is shown in the left bottom corner at the FWHM level of 8.17~mas.
The image supports the viewing angle being less than a half-opening angle even at large (kiloparsec, deprojected) spatial scales probed by the low-frequency observations.
Stokes~I image FITS file is is available at the CDS.
}
\label{f:L_image}
\end{figure}

We have processed 1.6~GHz VLBA observations of \pks, epoch 9~May~2017, experiment code BS257A. At ten times lower frequency, the VLBA is substantially more sensitive to distant optically thin jet emission and probes angular scales ten times larger than our MOJAVE stacked image. With a total on-source time of about 3~hr, these single epoch low-frequency observations reach a high dynamic range of about 7,000:1 and confirm the emission around the VLBI core, including the North-West direction (\autoref{f:L_image}). 
We note that our stacked image at 24~GHz also detects this emission (\autoref{f:1424_rmstack}).

\subsection{Rotation Measure at 15, 24, and 43 GHz}
\label{s:RM}

\begin{figure}
\centering
\includegraphics[width=1.03\columnwidth]{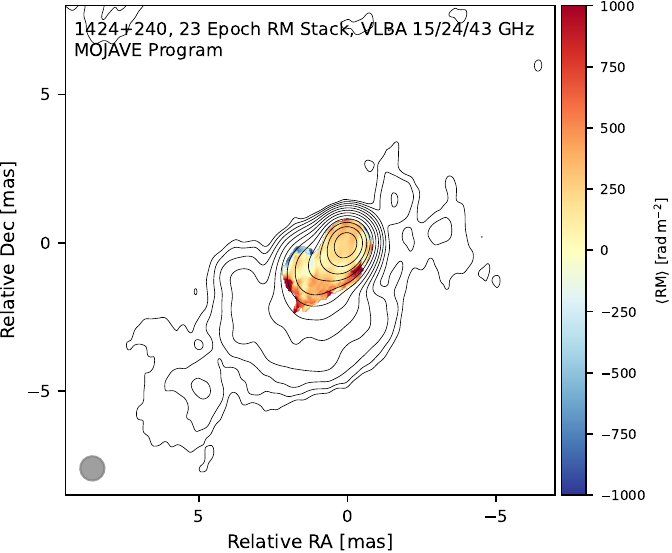}
\caption{Stacked mean RM map of 15/24/43~GHz VLBA data for \pks in the frame of the observer across 23 epochs from 15~August~2019 to 25~June~2021 in color scale. The contours show the mean 24~GHz Stokes~I; the first contour corresponds to $3\times$ the image rms level, with contours increasing by a factor of 2. The grey circle indicates the FWHM level of the circular restoring beam. The map shows the magnitude of RM is low and similar across the source. }
\label{f:1424_rmstack}
\end{figure}

Between August~2019 and June~2021, the MOJAVE monitoring conducted monthly 15~GHz, 24~GHz, and 43~GHz band polarimetric observations of 25 AGN (including \pks), totaling 23 epochs, in order to measure the rotation measure (RM; full results will be presented by Livingston and the MOJAVE team, in prep.). In \autoref{f:1424_rmstack} we show the mean RM of \pks across the two years of observations. This object is one of the lowest in total magnitude of observed RM for its core, with a median magnitude across all epochs of $302\pm108\,\mathrm{rad\,m^{-2}}$, compared to a population median magnitude of $1026\pm475\,\mathrm{rad\,m^{-2}}$. We also note that the median pixel-by-pixel standard deviation of RM for the core is $514\pm57\,\mathrm{rad\,m^{-2}}$, making it larger than the median magnitude of RM for the core. Interestingly, across the two years, the core of \pks does not vary greatly in the magnitude of RM, sitting in the bottom $\mathrm{25^{th}}$ percentile of time variability of sources. However, the jet of \pks has a time variability of RM similar to other sources within the sample. We also detect frequency dependent depolarization, which is typical of most sources within our sample. This combined with the low magnitude RM suggests that the Faraday rotation that we do see is external to the jet. 

As RM data is not available for all 42 epochs used in \autoref{f:1424stack} and individual epochs do not have RM data for the full jet region, we cannot apply a precise correction for the Faraday effect on EVPA. The median magnitude of RM for \pks corresponds to a EVPA rotation of $\sim6^{\circ}$, which is comparable to our EVPA errors from calibration and imaging.

The low magnitude and variability of RM especially close to the core of \pks are indicative of either a weak magnetic field or low thermal electron density along the line of sight. We may be seeing the `sweeping away' of the magneto-ionic medium, this would result in a reduction in the magnitude of RM. This effect would be more prominent within the center of the jet, while the edges would be less affected, which is supported by the magnitude and variability of RM being typical of other sources for the jet region of \pks. 

\end{appendix}

\end{document}